\documentclass{PoS}
\pdfoutput=1

\newcommand{\be}{\begin{equation}}
\newcommand{\ee}{\end{equation}}
\newcommand{\bea}{\begin{eqnarray}}
\newcommand{\eea}{\end{eqnarray}}
\newcommand{\beq}{\begin{equation}}
\newcommand{\eeq}{\end{equation}}
\newcommand{\ba}{\begin{array}}
\newcommand{\ea}{\end{array}}
\newcommand{\beqa}{\begin{eqnarray}}
\newcommand{\eeqa}{\end{eqnarray}}

\newcommand{\cA}{{\cal A}}
\newcommand{\cO}{{\cal O}}
\newcommand{\cB}{{\cal B}}
\newcommand{\BR}{{\cal B}}

\newcommand{\cL}{{\cal L}}

\newcommand{\no}{\nonumber}

\newcommand{\kpn}{K^+\to\pi^+\nu\bar\nu}
\newcommand{\klpn}{K_L\to\pi^0\nu\bar\nu}
\newcommand{\kno}{{K^0}}
\newcommand{\knb}{{\overline{K}^0}}
\newcommand{\Gno}{\Gamma_{K^0}}
\newcommand{\Gnb}{\Gamma_{\overline K^0}}
\newcommand{\Mno}{m_{K^{0}}}
\newcommand{\Mnb}{m_{\overline K^0}}
\newcommand{\GAML}{{\Gamma_L}}
\newcommand{\GAMS}{{\Gamma_S}}

\title{\boldmath KAON 2007: Conference Summary \unboldmath}

\ShortTitle{\boldmath Conference Summmary \unboldmath}

\author{\speaker{Gino Isidori}\thanks{This work
    is supported in part by by the  EU Contract No.~MRTN-CT-2006-035482, ``FLAVIAnet''.} \\
  INFN, Laboratori Nazionali di Frascati, Via E. Fermi 40, 
           I-00044 Frascati, Italy\\
  E-mail: \email{gino.isidori@lnf.infn.it}}

\abstract{A concise overview of the interesting recent developments 
           in kaon phyiscs discussed at KAON 2007 is presented.}

\FullConference{KAON International Conference\\
		May 21-25 2007\\	
	Laboratori Nazionali di Frascati dell'INFN, Rome, Italy}

\begin{document}

\section{Introduction}

There is no doubt that kaon physics has played  
a key role in our understanding of fundamental 
interactions in the last 60 years:
several corner stones of the  Standard Model (SM)
have been derived, directly or indirectly, from $K$ meson phenomenology.
What is less obvious, looking at this field from outside, 
is that kaon physics is still a very interesting subject. 
As I will try to summarise in the next five sections, 
the large number of interesting new results 
and stimulating discussions presented at this conference
provide a clear demonstration of this fact.

\section{\boldmath $V_{us}$ and the semileptonic challenge to the SM \unboldmath}
The  progress in the determination 
of $|V_{us}|$ and, more generally, of the precise SM tests
in $K_{\ell 2}$ and $K_{\ell 3}$ decays can be considered the main 
highlight of this conference. Before discussing 
the new experimental and theoretical results, it is worth spending few 
words about the interest of these measurements in general terms.
 
The determination of $|V_{us}|$ has now reached the $0.5\%$ level
of accuracy \cite{Palutan}. Thanks to the smallness of $|V_{ub}|$ and the 
high-precision of $|V_{ud}|$, this implies we can test at per-mil 
level the relation 
\begin{equation}
|V_{ud}|^2+|V_{us}|^2+|V_{ub}|^2 =1~.
\label{eq:unitarity}
\end{equation}
Beside testing the unitarity of the CKM matrix, 
which is naturally satisfied in the SM and in most of its 
extensions, the main interest hidden behind  Eq.~(\ref{eq:unitarity})
is a very stringent test of  the universality 
of weak interactions. As reminded us 
by Marciano~\cite{Marciano}, when extracting the $|V_{ij}|$ 
from a given $u_i \to d_j \ell \nu$ 
process, the corresponding rate is normalised to
$G_F^{(\mu)}$, or  the Fermi coupling determined 
from the muon decay. As a result, testing the 
unitarity relation in Eq.~(\ref{eq:unitarity}) 
is equivalent to testing the universality of 
weak interactions between quarks and leptons. 

Indeed, what is probed in low-energy experiment is not the 
gauge coupling of the $W$ to fermion fields
(which, by construction, is insensitive to new physics) 
but effective four-fermion interactions of the type
\beq
\cL^{\rm C.C.}_{\rm eff.} ~=~ G_F^{(ijkl)} ~ 
\bar u^i \Gamma d^j  ~ \bar\ell^k \Gamma \nu^l   {\rm ~+~h.c.}
\eeq
The values of the effective couplings $G_F^{ijkl}$ 
are not protected by gauge invariance and are potentially 
sensitive to physics beyond the SM. On general grounds \cite{AC}, 
one expects a power suppression of the non standard 
contributions of the type
\beq
G_F^{(ijkl)} = \left[G_F^{(ijkl)}\right]_{\rm SM} \left[ 1 + 
c \frac{M_W^2}{M_{\rm NP}^2} \right]~,
\eeq
where $M_{\rm NP}$ generically denotes the masses of the new particles.  
The coupling $c$ can be quite small if the new degrees of freedom 
appears only at the loop level [$c \sim ~ 1/(16\pi^2)$];
but the high precision reached in a few semileptonic $K$, 
$\pi$ and nuclear $\beta$ decays can partially 
compensate such suppression. For instance, comparing  $G_F^{(\mu)}$
with the effective coupling 
\beq
G_F^{\rm (CKM)} =  G_F \times [ |V_{ud}|^2+|V_{us}|^2 ]~,
\eeq
determined from  $K_{\ell 3}$ and nuclear $\beta$ decays,
allows us to exclude new heavy gauge bosons below $1$~TeV 
in specific beyond-SM frameworks~\cite{Marciano}.

A complementary class of weak universality tests is 
the comparison of the $|V_{us}|$ values, or better the effective weak couplings,
extracted from  $K_{e 3}$ (vector coupling only), 
$K_{\mu 3}$ (vector and scalar coupling), and $K_{\ell 2}$ 
(scalar coupling only). 
Several beyond-SM frameworks are severely constrained 
by such tests. For instance, right-handed currents which are 
naturally present in Higgs-less models can induce 
$\cO(1\%)$ 
violations of universality between $K_{e 3}$ and $K_{\mu 3}$ \cite{Stern}.
Similarly, in two-Higgs doublet models with large 
$\tan\beta$ ($\tan\beta =\langle H_U \rangle/\langle H_D \rangle$)
one expects violations of universality
between $K_{\mu 2}$ and $K_{e 3}$ around few$\times 0.1\%$ \cite{Paradisi}. 
Precise measurements of the decay distributions in  $K_{\ell 2}$ 
and $K_{\ell 3}$ decays could also constrain (or find evidences of) 
models with new light scalar particles, 
such as the $\nu$MSM~\cite{nuMSM}.

Last but not least, a particularly sensitive probe of new physics 
in semileptonic $K$ decays is the universality ratio 
\beq
R_K^{e/\mu} = \frac{ \cB(K\to e \nu) }{ \cB(K \to \mu \nu)}~.
\label{eq:Kme}
\eeq
This observable can be predicted to 0.04\% accuracy
within the SM~\cite{Cirigliano}, while its deviation from 
the SM can reach $\sim 1\%$ in realistic supersymmetric 
frameworks with new sources of lepton-flavour mixing~\cite{Paradisi}. 

In summary, the measurements of $K_{\ell 2}$ and $K_{\ell 3}$ 
decay parameters have reached a level o precision which allows us to 
perform new significant tests of the SM. These precision
tests are equally interesting and fully complementary
to the flavour-conserving electroweak precisions tests
and to the FCNC tests performed at $B$ factories. 

\subsection{On the theoretical description of $K_{\ell 2}$ and $K_{\ell 3}$ decays}

The high statistics on semileptonic $K$ decays 
accumulated in the last few years has motivated substantial 
progress in the theoretical description of these processes.  
In several observables the theoretical control is such that 
even the tiny deviations of lepton-flavor and/or quark-lepton universality 
mentioned above ($0.1\%$--$1\%$) can possibly be detected.
 
The two master formulae for the inclusive decay rates are
\beqa
\Gamma (K_{\ell 3 [\gamma] })  &=& C^0_{K_{\ell 3}}  I^{ K \ell} (  \lambda_i )  
\times | V_{us}~ f_{+}(0)|^2 \times \Big[ 
1 + 2\, \Delta^K_{SU(2)} + 2\,  \Delta^{K \ell}_{\rm EM} 
\Big]~, \qquad \\
\Gamma (K_{\ell 2 [\gamma] }) &=& C^0_{K_{\ell 2}}   m^2_\ell (m_K^2-m_\ell^2)
\times |V_{us}~f_K |^2 \times \Big[ 1 + 2 \Delta^{\ell}_{\rm EM} \Big]~.
\label{eq:Km2}
\eeqa
Here $C^0_K$ are overall coefficients which can be predicted 
with excellent accuracy (within the SM) in terms of $G_F^{(\mu)}$ 
and $\alpha_{\rm em}$, while the phase-space factors $I^{ K \ell}$
can be determined by experiments measuring the 
Dalitz plot slopes. The CHPT calculations of isospin-breaking and 
radiative-correction factors ($\Delta$'s) have recently been completed, 
cross-checked and scrutinized against variations of 
the input parameters~\cite{Cirigliano}. They are known with errors 
which range from a few~$\times 10^{-4}$ to a few~$\times 10^{-3}$,
with the largest uncertainty coming from the isospin-breaking 
term 
\beq
\Delta^K_{SU(2)} = (\Delta^{K^+}_{SU(2)} - \Delta^{K^0}_{SU(2)}) 
= (2.3 \pm 0.2)\times 10^{-2}~.
\label{eq:DS2th}
\eeq

The largest theoretical uncertainties (1.0-0.5\%)
are still confined in the two overall hadronic form factors,
$f_K$ and $f_{+}(0)$, whose best determinations are presently obtained 
from Lattice QCD. These uncertainties drop out in universality 
ratios, such as  $R_K^{e/\mu}$ in Eq.~(\ref{eq:Kme}). 
As a result, the precision in testing lepton-flavour universality 
or isospin-breaking effects is nowadays mainly an experimental issue. 
The limiting uncertainty in extracting $V_{us}$ and in
precision tests involving $f_K$ and $f_{+}(0)$ is still 
dominated by theory errors. However, the very recent progress 
of unquenched Lattice simulations (see sect.~\ref{sect:Vus}) 
allows to dream reaching $\cO(10^{-3})$ accuracies
also in these cases in a near future. 

\subsection{\boldmath  Experimental status of $K_{\ell 3}$ decays \unboldmath}
One of the virtues of the kaon system is that there is only a handful 
of leading decay modes, or branching ratios (BRs) above the $10^{-4}$ level.  
At fixed target experiments (KTeV, NA48 and ISTRA) the relative rates 
of these leading BRs are measured, while at KLOE, where the normalization of 
the kaon flux is know, direct measurements of the absolute branching ratios 
are also possible. The combination of all these data, together with life-time 
measurements, allows us to over-constrain the system. It is then possible 
to check the consistency of the various results and, possibly, to reduce 
the errors via global fits \cite{Flavia}.

\begin{figure}[t]
    \vskip -0.2 cm
\begin{center}
\includegraphics[width=0.2\textwidth]{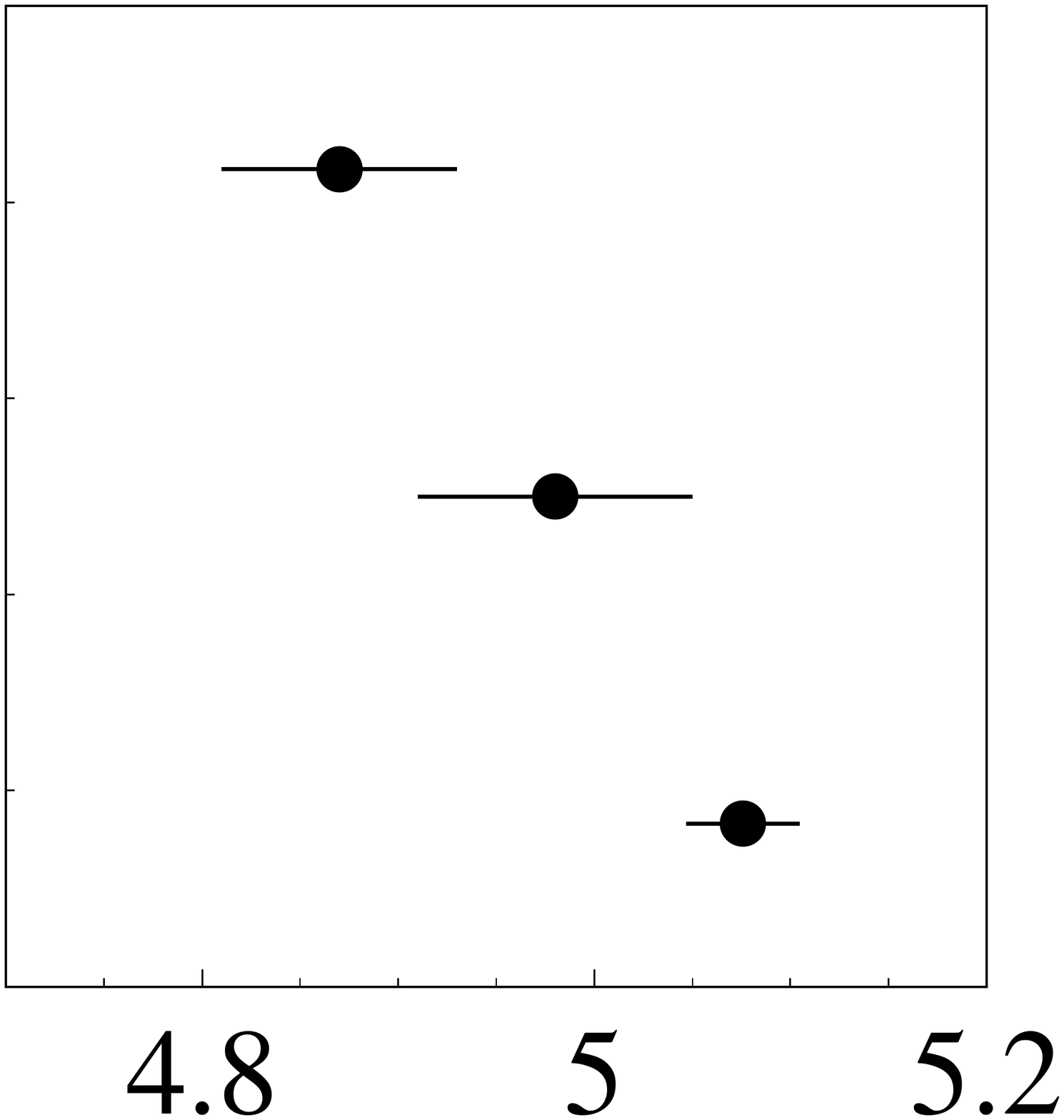}
\includegraphics[width=0.2\textwidth]{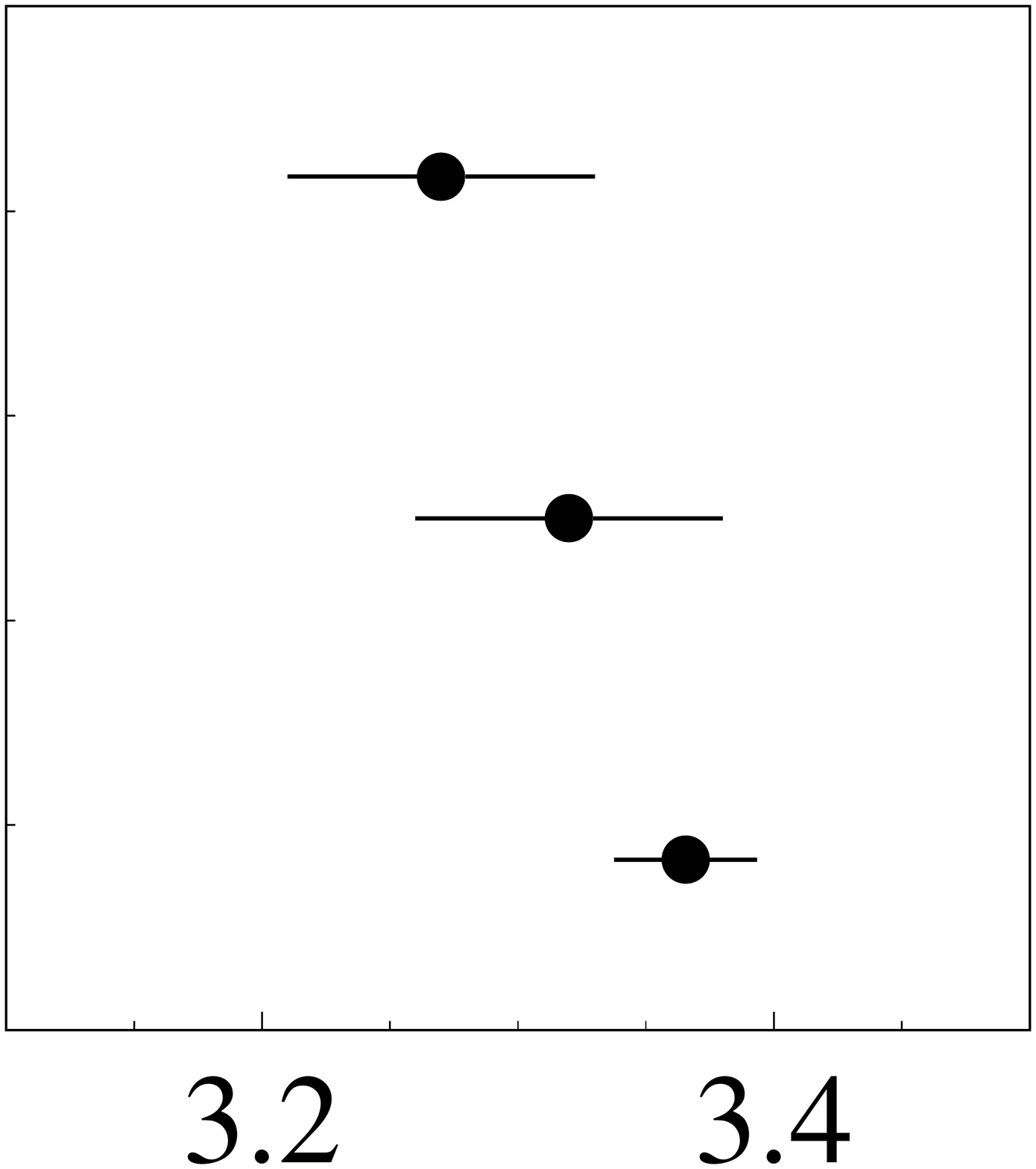}
\includegraphics[width=0.2\textwidth]{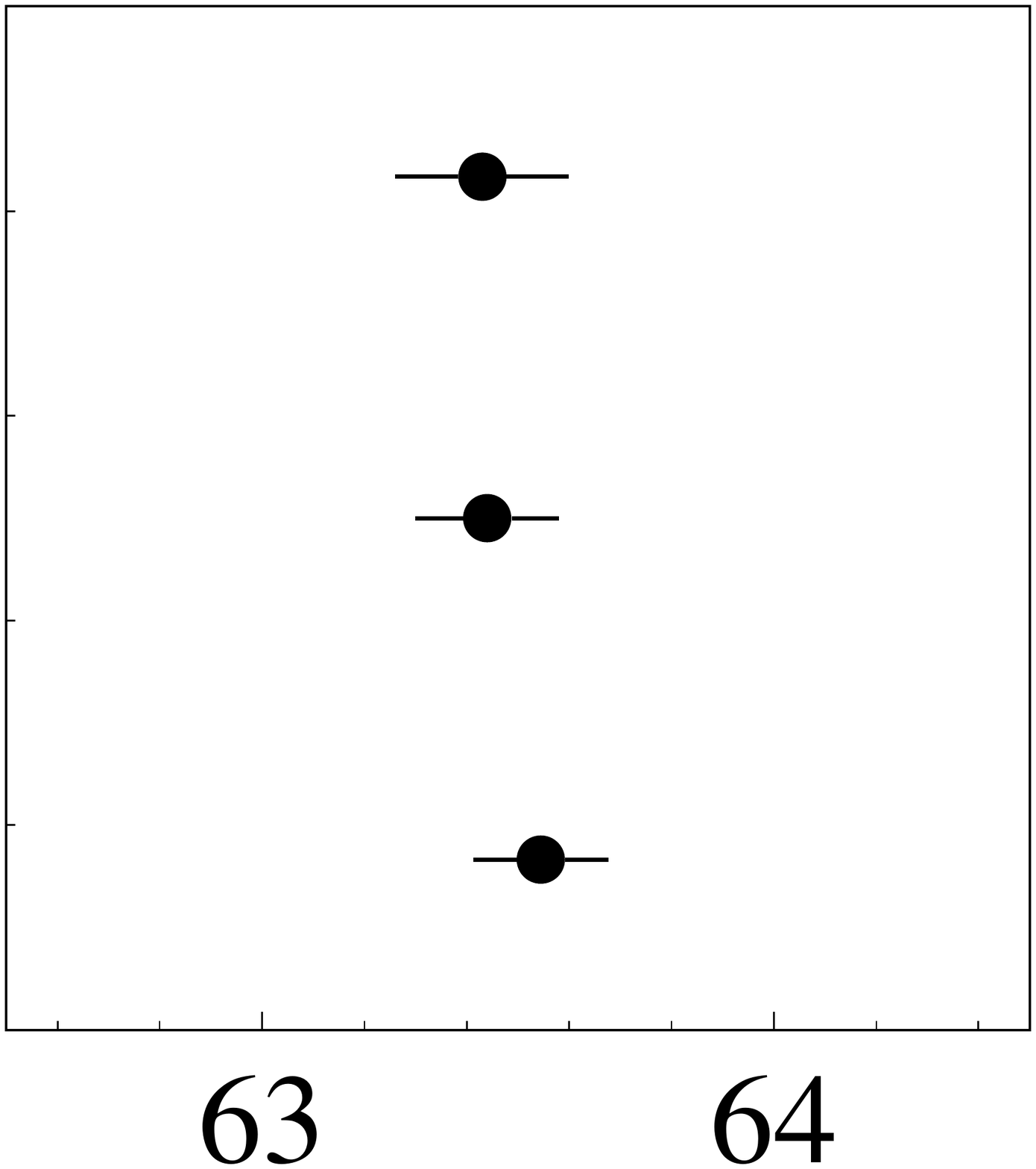}
\includegraphics[width=0.2\textwidth]{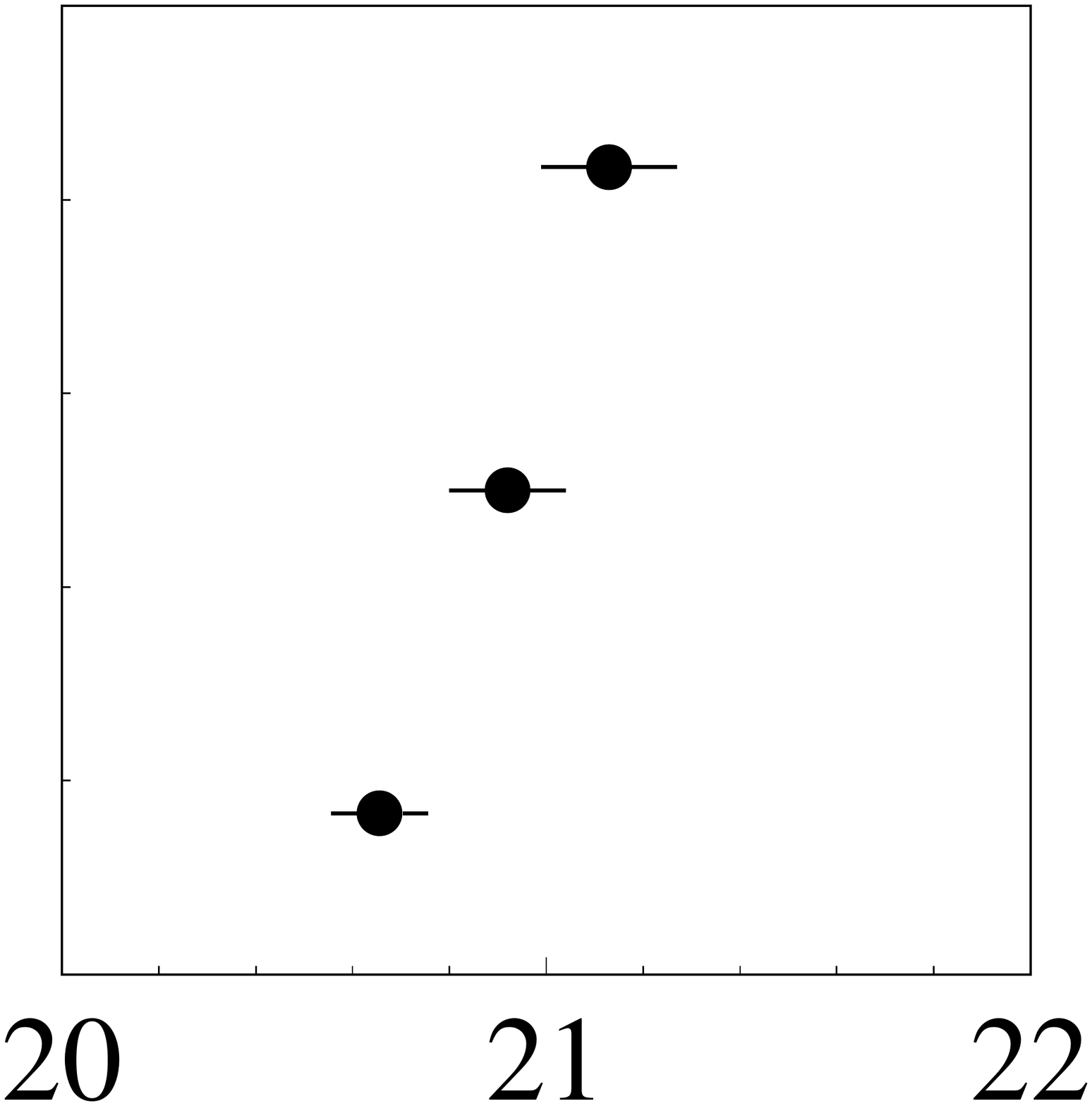}
\put(-330,90){$\BR(K^{\pm}_{e3}$) \% \hspace{1.3cm} 
$\BR(K^{\pm}_{\mu3})$ \% \hspace{1.3cm} 
$\BR(K^{\pm}_{\mu2})$ \% \hspace{1.3cm} 
$\BR(K^{\pm}_{2\pi})$ \% }
\put(-385,65){PDG '04 }
\put(-385,42){PDG '06 }
\put(-385,20){KAON '07}
\caption{Evolution in the measurements 
of some of the leading $K^\pm$ decay rates \cite{Palutan}.}
\label{fig:BRKp}
\end{center}
\end{figure}

In the $K_L$ system there is a good overall agreement 
of the results produced by the new generation of kaon 
experiments since more than one year. The situation is 
indeed almost unchanged since the latest PDG fit~\cite{PDG}. 
On the other hand, a significant improvement has occurred 
at this conference in the charged 
kaon system, thanks to new results announced by KLOE 
(absolute $\cB(K^+_{\ell3})$~\cite{Sciascia} and $\tau^+$~\cite{Massa}), 
NA48 ($\cB(K^+_{\ell3})/\cB(K^+_{\pi\pi})$~\cite{Dabro})
and ISTRA~($\cB(K^+_{\ell3}/\cB(K^+_{\pi\pi})$~\cite{Istra}).
As shown in Fig.~\ref{fig:BRKp}, 
there has been a substantial progress in the determination 
of the semileptonic branching ratios compared to
the PDG 2006 values.

Beside the improvement in the final determination of $V_{us}$,
the most interesting aspect of the progress on  $K^\pm$
semileptonic branching ratios is the direct determination 
of isospin-breaking effects from data~\cite{Palutan}:
\beq
\left(\Delta^K_{SU(2)}\right)^{\rm exp} 
= (2.86 \pm 0.38)\times 10^{-2}~.
\eeq
The good agreement of this result and the theoretical prediction in 
(\ref{eq:DS2th}) is a great success of the CHPT calculations of 
isospin-breaking effects~\cite{neufeld}. 
The consistency between theory and experiments in the leading 
isospin-breaking effects is reflected in the good consistency
of the fit to $|V_{us}| f_+(0)$ extracted from 
five different modes: $K^L_{e3}$, $K^L_{\mu3}$, $K^S_{e3}$, 
$K^{\pm}_{e3}$, $K^{\pm}_{\mu3}$. 
The new global fit yields 
\be
|V_{us}| f_+(0)  = 0.21663(47) \label{eq:vusf0}
\ee
with $\chi^2/N_{\rm dof} =2.62/4$ (Prob$ = 62\%$). 

Despite the remarkable progress since 2004 (when CKM unitarity 
seemed to be violated at the $3\sigma$ level\ldots),
it should stressed that we are still far from an ideal situation
as far the experimental results are concerned. In the $K^{\pm}$ 
system not all the leading decay modes have been re-measured. 
More important, the recent measurements of slope and curvature 
of the form factors --especially the scalar slope-- are not in good 
agreement~\cite{Franzini}. 
This inconsistency has a minor direct impact in Eq.~(\ref{eq:vusf0}); 
however, it prevent us to perform interesting chiral tests 
(see~Ref.~\cite{Stern,Bijnens}) 
of the methods used to estimate $f_+(0)$ (including Lattice QCD).
There is certainly room for significant improvements in view 
of KAON 2009! 

\subsection{\boldmath  Lattice progress on $f_+(0)$ and $f_K/f_\pi$ 
and the extraction of $V_{us}$ \unboldmath}
\label{sect:Vus}

\begin{figure}[t]
    \vskip -5.5 cm 
    \begin{center}
    \begin{minipage}[t]{0.48\linewidth}
    \includegraphics[width=1.2\linewidth]{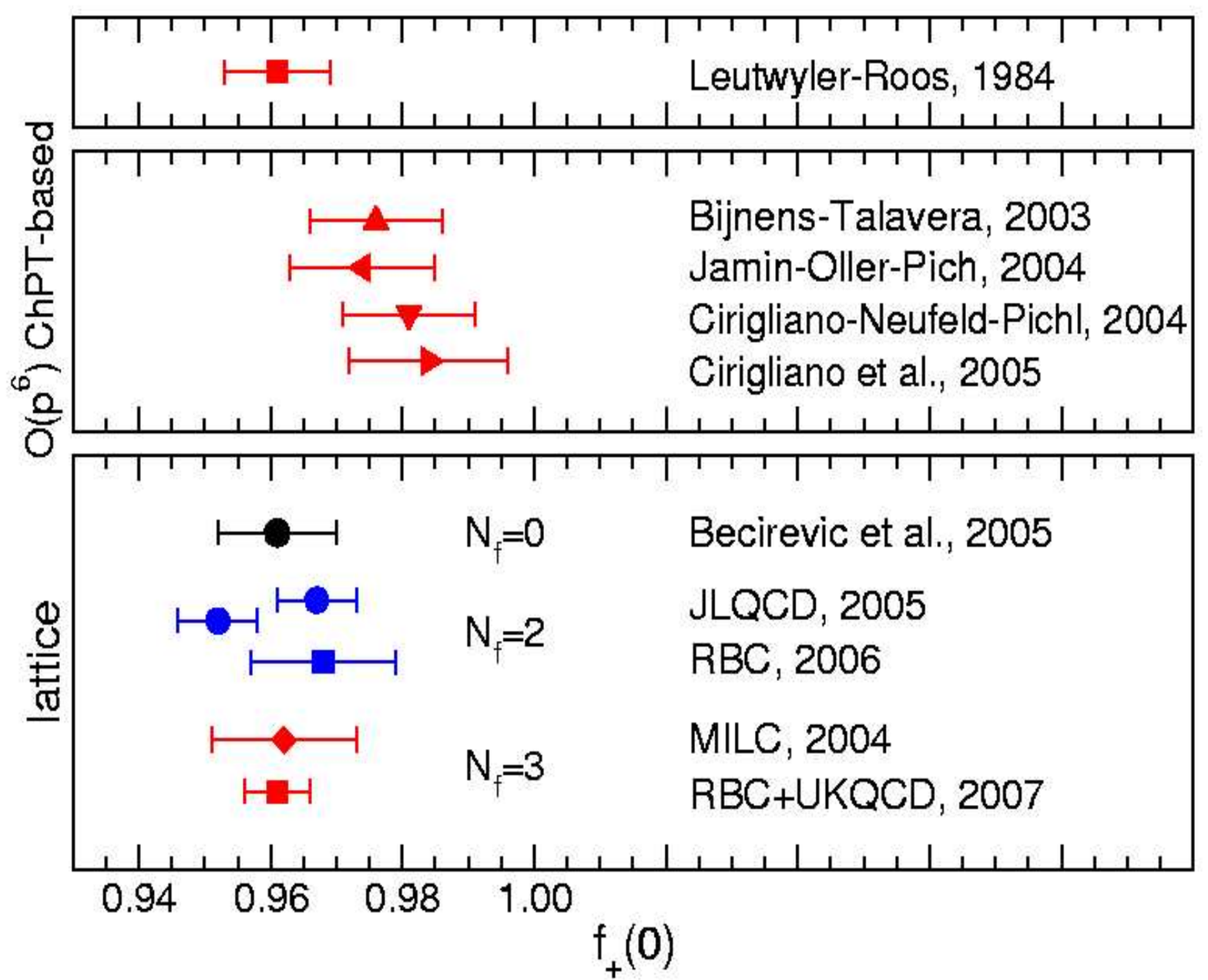}
     \caption{Comparison of analytic and lattice results for $f_+(0)$ \cite{Kaneko}
($N_f$ denote the number of dynamical fermions).\label{fig:lat1}}
    \end{minipage}
    \hspace*{\fill}
    \begin{minipage}[t]{0.48\linewidth}
    \includegraphics[width=1.2\linewidth]{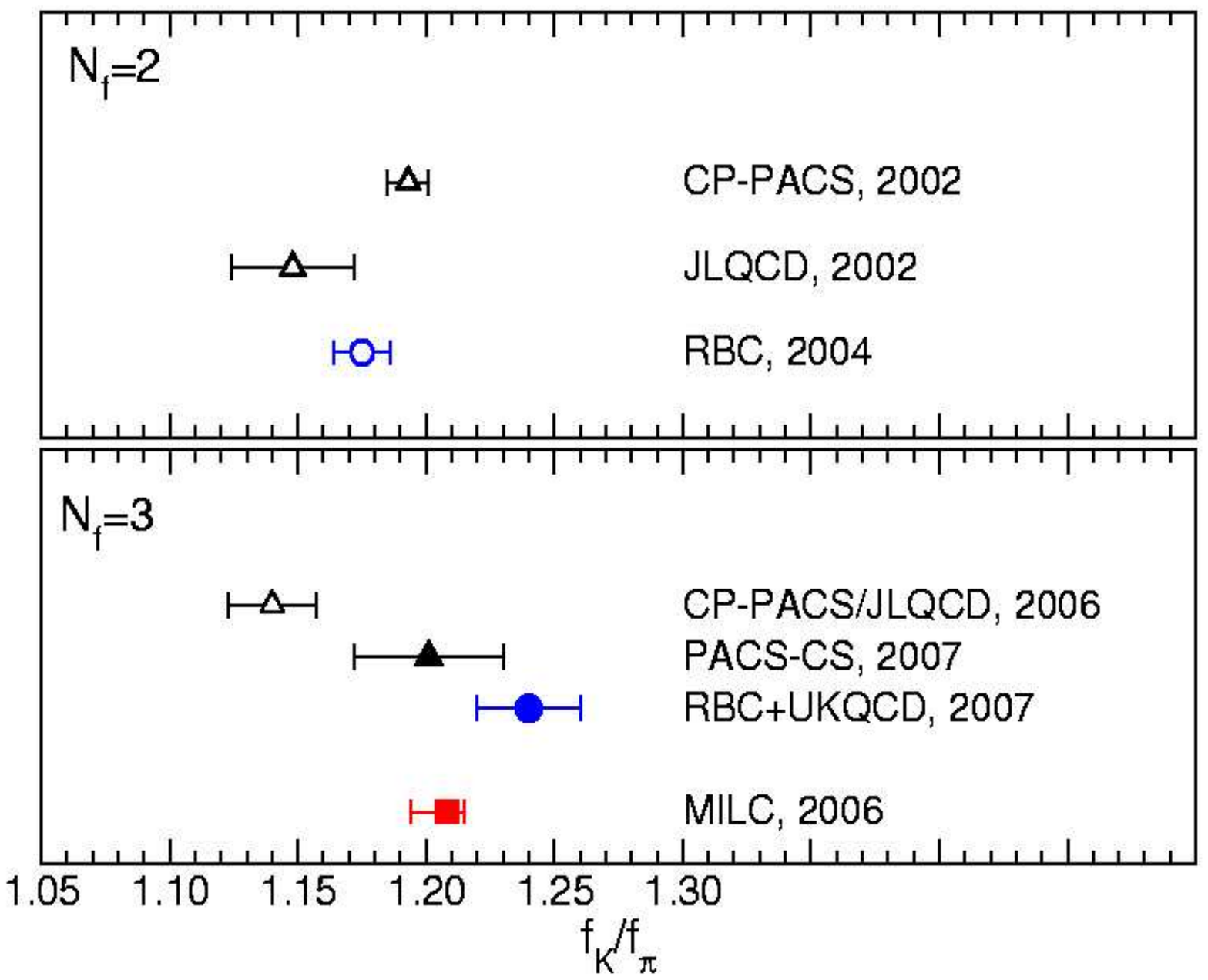}
     \caption{Comparison of recent unquenched and partially-quenched 
 lattice results for $f_K/f_\pi$ \cite{Kaneko}.\label{fig:lat2}}
    \end{minipage}
   \end{center}
\end{figure}

As nicely summarised by Kaneko~\cite{Kaneko} and Juttner~\cite{Jutt},
in the last two years 
there has been a dramatic improvement in the quality of lattice simulations 
relevant to kaon form factors. 
The two main obstacles which so far prevented reliable estimates, namely 
the quenched approximation and the chiral extrapolation, have essentially
been removed. The most recent simulations are performed with 2 or 3 dynamical
fermions and with pion masses around or below $\approx 300$~MeV, 
where the chiral extrapolation does not represent a severe problem. 

The recent results for $f_+(0)$ and $f_K/f_\pi$ are shown in Fig.~\ref{fig:lat1} 
and~\ref{fig:lat2}. In the case of $f_+(0)$ there is still some competition
from analytical approaches, while in the case of $f_K/f_\pi$ only lattice results 
are able to reach the $\cO(1\%)$ level. This of course does not mean that
analytical calculations of kaon form factors are useless.  
The non-trivial effort of computing these form factors in CHPT at one or 
two-loop level~\cite{Bijnens} is an essential tool to improve and 
control the chiral extrapolation of the recent lattice results. 

As far as $f_+(0)$ is concerned, it is interesting to note that all lattice 
results, and especially the most recent and precise one by Ref.~\cite{Antonio}, 
are in excellent agreement with the old result by Leutwyler and Roos. 
At first sight this is somehow surprising given the large $\cO(p^6)$ 
contributions appearing in CHPT at the two-loop level~\cite{BT} 
(which are responsible for the large central values of $f_+(0)$ 
obtained in CHPT + large $N_c$ approaches~\cite{largeN}). 
On the other hand, it should be stressed that the $\cO(p^6)$ 
CHPT calculation of this quantity is not complete, given the absence 
of a model-independent information on the corresponding 
counterterms. If confirmed by a more reliable chiral extrapolation,
the result of Ref.~\cite{Antonio} would then provide a very
useful insight about CHPT beyond the one-loop level. 

\begin{figure}[t]
    \vskip -2.5 cm
\begin{center}
\includegraphics[width=0.6\textwidth]{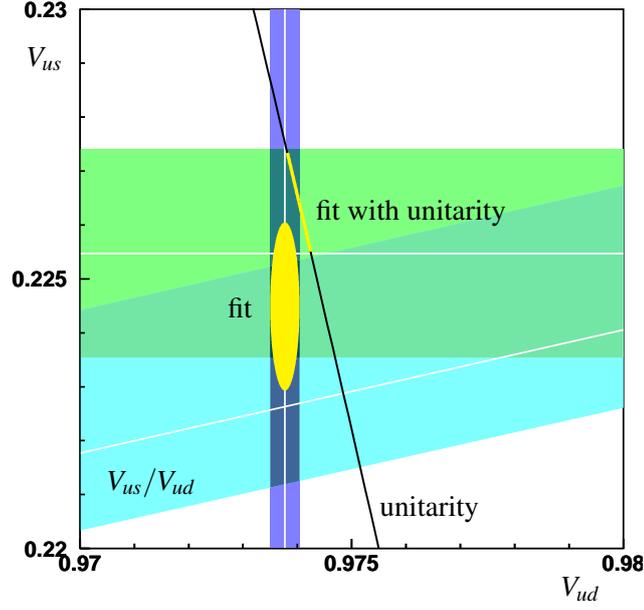}
\put(-230,230){$V_{us}$}
\put(-30,30){$V_{ud}$}
\put(-200,70){$V_{us}/V_{ud}$}
\put(-155,135){fit}
\put(-122,172){fit with unitarity}
\put(-98,60){unitarity}
    \vskip -0.5 cm
\caption{Results of the recent fits to $|V_{us}|$ (from $K_{\ell 3}$)
and  $|V_{us}|/|V_{ud}|$ (from $K_{\ell 2}$) compared to the world
average value of $|V_{ud}|$.
\label{fig:vusvud} }
\end{center}
\end{figure}

The present global fit to $|V_{ud}|$ and $|V_{us}|$ using the 
Leutwyler-Roos result for $f_+(0)$ and $f_K/f_\pi = 1.208(2)(^{+7}_{-14})$~\cite{milc}
--used to determine $|V_{us}|/|V_{ud}|$ from $\Gamma(K_{\ell 2})/\Gamma(\pi_{\ell 2})$--
is shown in Fig.~\ref{fig:vusvud}. As can be noted, the agreement with 
CKM unitarity is excellent.

All the recent lattice results on kaon form factors 
at the  $\cO(1\%)$ level are still based 
on a limited set of quark masses (especially 
in the case of $f_+(0)$) and/or specific actions (especially 
in the case $f_K/f_\pi$). Therefore the quoted errors 
should still be taken with some care. Nonetheless it is clear 
that we are entering a new era, where lattice will help us both
to test the SM at the per-mil level and to shed light on the 
convergence of the chiral expansion beyond the one-loop level.

\subsection{\boldmath The $R_K^{e/\mu}$ ratio \unboldmath}

As anticipated, the ratio  in Eq.~(\ref{eq:Kme}) can be computed 
to an excellent accuracy within the SM and provides an interesting 
probe of non-standard scenarios which violate lepton flavour universality.  
The recent work of Ref.~\cite{bib:masiero}, which has pointed out 
the possibility of deviations from the SM 
of up to $\sim 1\%$ in realistic supersymmetric scenarios, 
has triggered a renewed interest in the  precise measurements
of this ratio~\cite{Wanke}.

\begin{figure}
    \vskip -1.0 cm
    \begin{center}
    \begin{minipage}[t]{0.48\linewidth}
    \raisebox{48mm}{      \begin{tabular}{lc}
      
        \hline \hline
                                                  & $R_K^{e/\mu}$ $[10^{-5}]$  \\ \hline
        PDG 2006~\cite{PDG}                       & $2.45 \pm 0.11$ \\
        NA48/2~'03~prel.~\cite{bib:Ke2_2003}   & $2.416 \pm 0.043 \pm 0.024$ \\
        NA48/2~'04~prel.~\cite{bib:Ke2_2004}   & $2.455 \pm 0.045 \pm 0.041$ \\
        KLOE prel.~\cite{bib:Ke2_KLOE}            & $2.55 \pm 0.05 \pm 0.05$ \\ \hline
        SM prediction                             & $2.472 \pm 0.001$ \\
        \hline \hline
      \end{tabular}}
     \vskip -2.5 cm 
     \caption{Current experimental data on $R_K^{e/\mu}$ (left).
             Exclusion limits at $95\%$ CL in the 
             $\tan \beta$--$M_{H^\pm}$  plane from  $R_K^{e/\mu}$ 
             for different values of the LFV coupling 
             $\Delta_{13}$~\cite{bib:masiero}. For comparison,  
	     the present limits from $B \to \tau \nu_\tau$ decays are also shown.
     \label{fig:susylimit} }
    \end{minipage}
    \hspace*{\fill}
    \begin{minipage}[t]{0.48\linewidth}
    \includegraphics[width=0.90\linewidth]{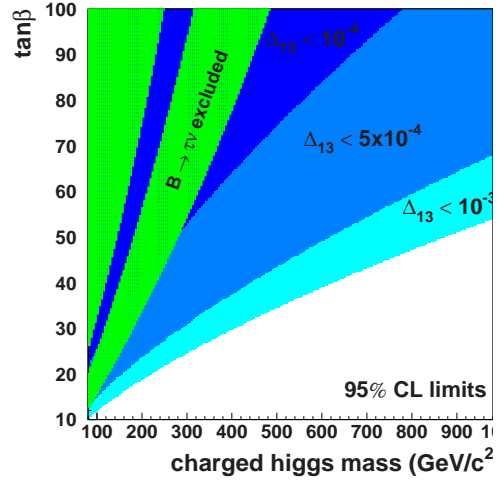}
    \end{minipage}
   \end{center}
\end{figure}

The limiting factor in the determination $R_K^{e/\mu}$ is the $K\to e\nu$ rate, 
whose experimental knowledge has been quite poor so far. The current official
world average, which implies $R_K^{e/\mu} = (2.45 \pm 0.11) \times 10^{-5}$ dates 
back to experiments from the 70's~\cite{PDG}. The situation has finally 
changed thanks to a series of preliminary results by NA48/2 and KLOE
(see Fig.~\ref{fig:susylimit}), two of which have been announced at this conference.
The two results by NA48/2, being based on different data sets (2003
and 2004, respectively) with different running conditions, 
should be regarded as completely independent. 
Combining these new results with the PDG value yields~\cite{Wanke}
\be
 R_K^{e/\mu} = ( 2.457 \pm 0.032 ) \times 10^{-5}~.
\label{eqn:ke2kmu2}
\ee
This result  is in good agreement with the SM expectation 
and has a relative error ($\sim 1.3\%$) three times smaller 
compared to the previous world average.

In the supersymmetric framework of Ref.~\cite{bib:masiero} this 
result implies significant constraints in the $\tan \beta$--$M_{H^\pm}$ plane
(see Fig.~\ref{fig:susylimit}). The constraints depend on the 
lepton flavour violating (LFV) coupling 
$\Delta_{13}$, whose natural values lies below $10^{-3}$
($\Delta_{13} \sim \delta^\ell_{RR} \times \alpha/(2\pi)$).
It is interesting to note that even for 
$\Delta_{13} \sim 10^{-4}$ the bounds extracted from 
$R_K^{e/\mu}$ are competitive with  the LFV-independent 
bounds presently derived from $B \to \tau \nu_\tau$. 

Further improvements in the knowledge of $R_K^{e/\mu}$ are 
foreseen in the near future. The preliminary KLOE measurement
has a conservative estimate of the systematic uncertainty
and is not based on the full statistics. Adding the remaining 
statistics and improving the reconstruction efficiency 
should reduce its error down to the 1\% level. 
An error on $R_K^{e/\mu}$ of about $0.3\%$ 
is the ambitious goal of the 2007 dedicated run of the CERN-P326 
collaboration (the successor of NA48)~\cite{bib:Ke2_2004,Ruggiero}
If these expectations will be fulfilled, in a short term the 
world average of $R_K^{e/\mu}$ will decrease by an additional
factor of four. This could open the possibility of finding clear
deviations from the SM, or to set very stringent limits in the 
parameter space of realistic supersymmetric models.

\section{\boldmath  The {\em mystic world} of $\pi\pi$ phase shifts  \unboldmath}

While $K_{\ell 2}$ and $K_{\ell 3}$ decays represent a unique 
observatory on electroweak interactions, 
$K_{3\pi}$ and $K_{\ell 4}$ decays provide a privileged 
observatory on strong interactions. The key observables 
here are the $\pi\pi$  phase shifts near threshold,
and particularly the two $S$-wave $\pi\pi$ 
scattering lengths. These hadronic 
observables can be predicted with an excellent 
accuracy in CHPT~\cite{CGL}
\bea
a_{0}~m_{\pi^+} &=& 0.220 \pm 0.005~, \qquad 
a_{0}~m_{\pi^+}~=~ -0.044 \pm 0.001~, \no \\
(a_{0}- a_{2})~m_{\pi^+} &=& 0.265 \pm 0.004~.
\label{eq:aIchpt}
\eea
The above results are direct consequences 
of the basic CHPT assumptions about the breaking 
of chiral symmetry in the QCD vacuum.
Confronting these predictions with experiments 
is therefore a way to test and improve our knowledge 
about the QCD vacuum~\cite{Colangelo}.

Till very recently the experiments where unable to match 
the high precision of the predictions in Eq.~(\ref{eq:aIchpt}).
The situation has substantially improved mainly
thanks to the high-precision measurements 
of $K^{+}_{3\pi}$ and $K^{+}_{\ell 4}$ decay 
distributions performed by NA48/2.
However, the situation is far from being completely understood.
The high accuracy of the new experimental results 
poses new challenging questions to the theoreticians~\cite{Juerg}: 
how to relate the predictions in Eq.~(\ref{eq:aIchpt}),
which have been obtained in an ideal framework of perfect 
isospin symmetry ({\em the paradise}), to what is measured 
in experiments ({\em the real world}), where we cannot 
switch-off the isospin-violating electromagnetic interactions.

\subsection{\boldmath  The $K\to 3 \pi$ cusp \unboldmath}

\begin{figure}[t]
\begin{center}
   \includegraphics[width=1.0\linewidth]{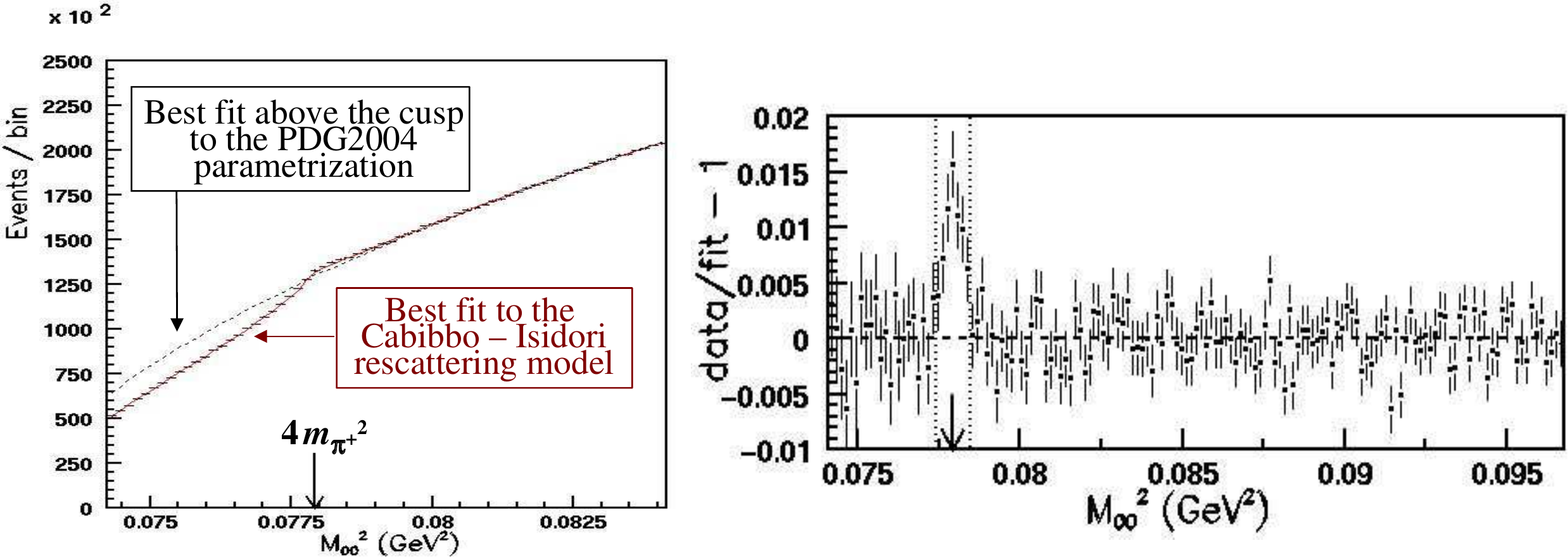}
\end{center} 
\caption{$M_{\pi^{0}\pi^{0}}$ spectrum 
in $K^{+}\rightarrow\pi^{+}\pi^{0}\pi^{0}$, close to the 
cusp region (left), compared to the PDG parameterization 
(no singularity) and the model of Ref.~\cite{CI}.
Deviation of the fitted spectrum (using the parameterization 
of Ref.~\cite{CI}) and data (pulls) with statistical errors 
(NA48/2 combined 2003+2004 data set)~\cite{Dilella}. \label{fig:cusp}}
\end{figure}

As pointed out by Cabibbo in 2004~\cite{Cabibbo:2004gq}, 
the rescattering of the final state 
pions produces a prominent cusp in the $M_{\pi^{0}\pi^{0}}$
spectrum of the $K^{+}\rightarrow\pi^{+}\pi^{0}\pi^{0}$ decay.
This effect can be used to obtain a precise determination 
of the $a_I$ and, particularly, of the $a_{0}-a_{2}$ combination. 
The NA48/2 collaboration has just released~\cite{Dilella} a combined 
fit of their 2003+2004 data set using the parameterization 
of Ref.~\cite{CI} (see Fig.~\ref{fig:cusp}).

The approach of  Ref.~\cite{CI} is based on a systematic 
expansion of the singular terms of the  $M_{\pi^{0}\pi^{0}}$
distribution in powers of $\pi\pi$ scattering lengths: 
a framework which is more efficient (as far as the extraction 
of the  $a_I$ is concerned) and substantially 
simpler than ordinary CHPT. However, this method allows us to 
take into account only the isospin-breaking effects which 
can be re-absorbed into the definition of pion masses and 
scattering lengths (what we can call {\em factorisable} radiative corrections).
As can be noted, the agreement between data and theory is 
good but for a narrow region around the cusp,
where the formation of the pionium occurs.
Cutting out this region, 
the results of the fit are~\cite{Dilella}
\begin{displaymath}
\begin{array}{rcrcrcrcr}
(a_0-a_2) &=&0.261&\pm&0.006_{\rm stat.}&\pm&0.003_{\rm syst.}&\pm&0.001_{\rm ext.}\\
a_2 &=&-0.037&\pm&0.013_{\rm stat.}&\pm&0.009_{\rm
syst.}&\pm&0.002_{\rm ext.}
\end{array}
\end{displaymath}
where the theoretical error induced by the parameterization 
of Ref.~\cite{CI} (estimated to be around $5\%$) is not included.  
The small statistical error, the nice agreement with 
the theory predictions in Eq.~(\ref{eq:aIchpt}), 
and the high quality of the global fit 
to the $M_{\pi^{0}\pi^{0}}$ distribution is quite impressive!

To fully exploit the high potential of this method (and these 
excellent data) a further improvement to reduce the theory error is need. The
recent works in Ref.~\cite{Bern,Ximo} are a useful steps forward in 
this direction: they both confirm (at least numerically) the 
parameterization of Ref.~\cite{CI} around the cusp up in absence 
of non-factorisable electromagnetic corrections. 
Moreover, the tiny numerical difference of the three methods suggest 
that the $\cO(a_I^3)$ terms are negligible. However, the key issue, 
namely the estimate of the non-factorisable radiative corrections,
has not been addressed yet. The approach of the Bern group~\cite{Bern}, 
based on a non-relativistic field theory, provide a systematic 
tool to evaluate these effects. 

Last but not least, it is worth 
to stress that a further reduction of both theory and experimental 
errors on the $a_I$ extracted from this method requires
a precise re-measurement of the $K^+ \to \pi^+ \pi^+ \pi^-$ channel~\cite{Juerg}. 
Indeed the leading singularity in the $M_{\pi^0 \pi^0}$ distribution 
arises by the $\pi^+ \pi^- \to \pi^0\pi^0 $ scattering, whose 
strength (compared to the regular terms) is controlled by 
the ratio $\Gamma(K^+ \to \pi^+ \pi^+\pi^-)/\Gamma(K^+ \to \pi^+ \pi^0\pi^0)$ .

\subsection{\boldmath  $K_{\ell 4}$  decays and global
fit to  the $a_I$ \unboldmath}

Historically  $K_{\ell 4}$  decays have been the main source of 
information on $\pi\pi$ phase shifts. Here the four-body kinematics 
allows different isospin and angular-momentum states for 
the two-pions, which can be disentangled studying their 
angular distribution. Contrary to the cusp effect just discussed, 
the phase shifts are measured above threshold and 
the extraction of the scattering lengths is performed 
with the $s_{\pi\pi} \to 4m_\pi^2$ extrapolation.

\begin{figure}[t]
\begin{center}
   \includegraphics[width=0.5\linewidth,angle=-90]{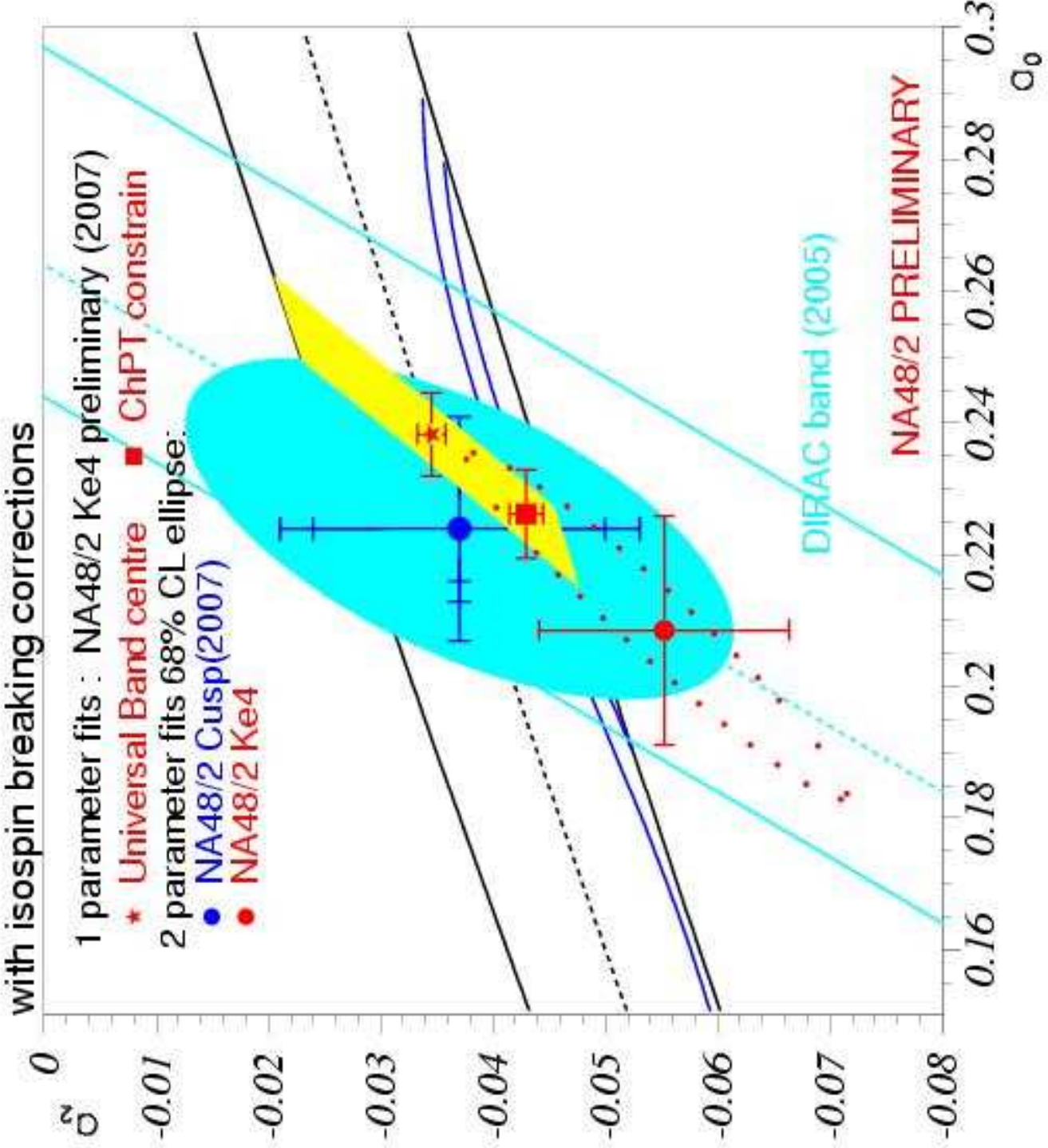}
\end{center} 
\caption{Recent NA48/2 results on the $a_0$ 
 and $a_2$~\cite{Dilella}. \label{fig:aiglob}}
\end{figure}

The first high-statistics analysis of the scattering lengths in 
$K_{\ell 4}$ has been performed a few years ago by the BNL-E85~\cite{Pislak:2003sv} 
collaboration. Their results are in good 
agreement with the CHPT predictions in Eq.~(\ref{eq:aIchpt}), 
although the final error is still quite large ($\approx 6\%$ 
on $a_0$, fixing the $a_2$--$a_0$ correlation form theory). NA48/2 has just released 
a preliminary analysis of their 2003 data set which 
has already twice the statistic of the BNL-E85 experiment~\cite{Bloch},
and their full data set (2003+2004) could allow to 
further double it.  
A final value for the $a_I$ has not been presented yet, 
but some interesting conclusions can already be drawn.
Isospin-breaking effects in the phase shifts, which have been 
ignored in previous analyses, turn out to be quite relevant. 
Only after including these effects (which are still 
under theoretical investigation by Gasser and  
collaborators~\cite{Juerg}), the $a_I$ extracted form 
$K_{\ell 4}$ are consistent with those derived from 
the cusp analysis (and with the theoretical predictions). 
The $a_0$ and $a_2$ values extracted from the 
$K_{\ell 4}$ analysis have a mild correlation, 
different from the one arising from the cusp effect
(see Fig.~\ref{fig:aiglob}). Imposing 
external theoretical constrains on the  $a_0$--$a_2$ 
plane, as for instance done in Ref.~\cite{Pislak:2003sv}, 
leads to sizable variations in the output results. 
On the other hand, the 
different correlation between cusp and $K_{\ell 4}$
data implies that combining these two data set  
we could finally challenge the theoretical predictions
for the $a_I$ in a model-independent way.

\section{\boldmath  The {\em chaotic world} of non-leptonic 
and radiative decays \unboldmath}

The theoretical description of branching ratios 
and CP violating asymmetries in non-leptonic $K$ decays 
is very different from the pristine world of semileptonic 
$K$ decays and $\pi\pi$ scattering. Because of the complicated 
interplay of strong and weak interactions, theory is still 
well behind the high-precision level reached by experiments.  

As discussed by C.~Sachrajda and R.~Mawhinney ~\cite{Sachrajda}, 
in the long run there is a realistic hope that lattice QCD 
calculations of non-leptonic channels will  
get close to the precision reached in experiments. 
All conceptual problems involved in a full QCD calculation 
of $K\to 2\pi$ amplitudes (in particular the delicate 
issue of final-state interactions) have been solved. 
What remain to be faced are practical problems:
the $K\to 2\pi$ amplitude require the evaluation of 
4-point Green functions of several operators 
(compared to the 2- and 3-point functions of a
single operator involved in semileptonic form factors); 
an excellent control of chiral corrections is needed,
and the control of final-state interactions require 
large volumes. At present, all this make impossible 
to directly confront theory and experiments in $K\to 2\pi$,
especially in delicate quantities such as $\epsilon^\prime/\epsilon$.
But the situation is likely to improve in the future
with larger CPU and improved algorithms. 

In the meanwhile, experiments are producing new interesting results.
The NA48/2 collaboration has announced the final results on the 
$K^{\pm}\to (3\pi)^\pm$ Dalitz-plot asymmetries~\cite{:2007yfa}:
\bea
A^c_g &=& \frac{g^+ - g^-}{g^+ + g^-}  = (-1.5 \pm 2.1) \times 10^{-4}
\qquad [K^{\pm}\to\pi^\pm\pi^+\pi^-]~, \no \\
A^n_g &=& (1.8  \pm 1.8) \times 10^{-4} 
\qquad [K^{\pm}\to\pi^\pm\pi^0\pi^0]~,
\eea
which show no evidence of CP violation at the $10^{-4}$ level. 
This result, which is expected within the SM by general 
arguments~\cite{DIP} (and confirmed by a detailed 
CHPT analysis~\cite{Prades,Gamiz}), allows to exclude 
some exotic NP scenarios~\cite{DIM}.

\begin{figure}[t]
\begin{center}
   \includegraphics[width=0.35\linewidth,angle=90]{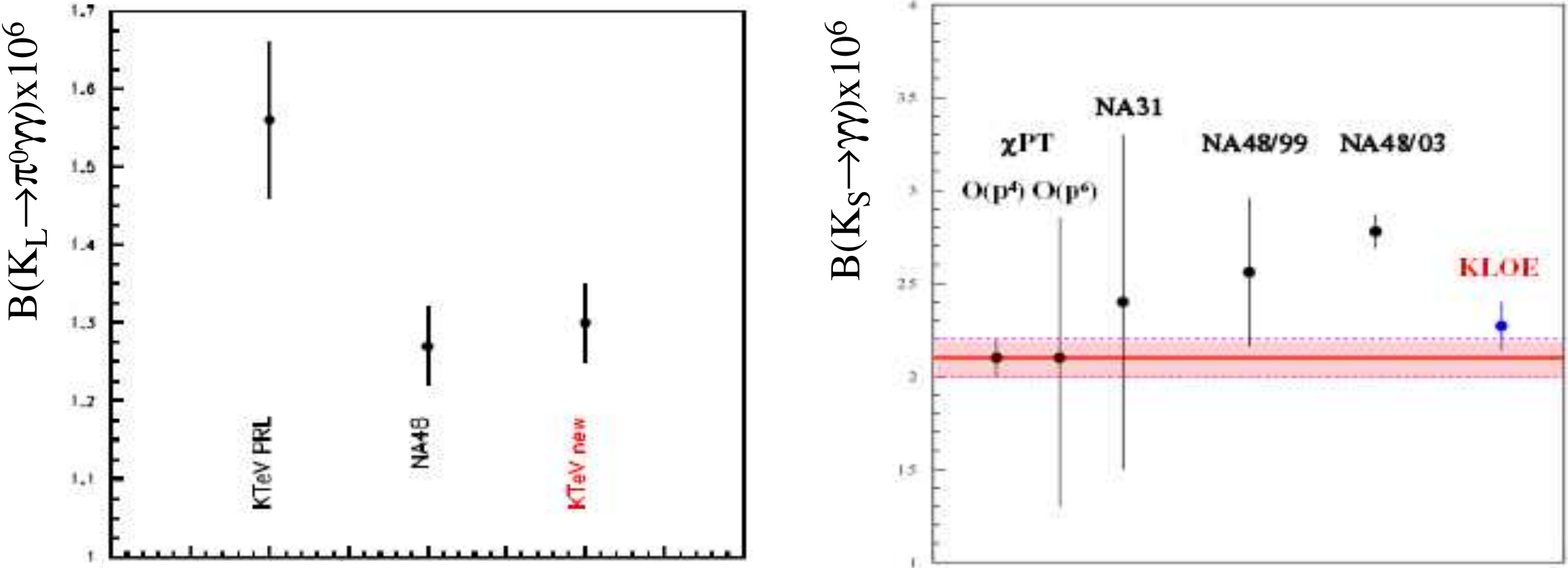}
\end{center} 
\caption{Recent results on two-photon modes: 
$\BR(K_L \to \pi^0 \gamma \gamma )$ from KTeV~\cite{KTeV1} and 
$\BR(K_S \to \gamma \gamma )$ from KLOE~\cite{Martini}.
 \label{radiativi}}
\end{figure}

The situation of radiative non-leptonic $K$ decays is 
somewhat better than the pure non-leptonic sector since CHPT
allows us to derive non-trivial relations~\cite{Prades}.
We cannot test the electroweak sector of the SM, but we can understand better the 
interplay of strong and weak interactions a low energies.  
Particularly interesting are the two-photon modes
$K_S \to \gamma \gamma$  and $K_L \to \to \pi^0 \gamma \gamma$.
The rates of these channels are unambiguously predicted 
in CHPT at $\cO(p^4)$ and their precise measurements 
allow us to explore the largely unknown sector of CHPT at 
$\cO(p^6)$. Here we have seen two interesting new experimental results
(see Fig.~\ref{radiativi}). 
A new measurement of $\BR(K_L \to \pi^0 \gamma \gamma )$
by KTeV~\cite{KTeV1} has finally 
clarified the long-standing discrepancy with NA48 
(incidentally, this is a also a very good news for $K_L \to \pi^0 e^+ e^+$, 
since it implies that its CPC component is totally negligible.~\cite{BDI}). 
On the other hand, KLOE has announced a new measurement of the 
$K_S \to \gamma \gamma$ rate~\cite{Martini} which is barely consistent with 
a previous NA48 result. A clarification of the $K_S \to \gamma \gamma$
result would be useful to estimate the real size of the $\cO(p^6)$ terms
in channels which are free from vector-meson enhanced contributions. 

\section{\boldmath  Challenging the basis: CPT and QM tests \unboldmath}

The neutral kaon system offers a unique 
possibility to challenge some of the foundations of our 
description of fundamental interactions,
such as CPT invariance and Quantum Coherence. 
Exact CPT invariance is expected in any quantum field theory respecting 
the general hypotheses of Lorentz invariance, locality and unitarity. 
However, these hypotheses are likely to be violated 
at very high energy scales, where the quantum effects of gravitational 
interactions cannot be ignored. As shown by Mavromatos~\cite{Mavromatos},
CPT non-invariance naturally arises in quantum-gravity frameworks 
as a consequences of Lorentz violations and, in specific 
frameworks, the CPT symmetry 
could even be hill-defined. On the phenomenological side, 
since we still miss a consistent theory of quantum gravity,
it is hard to predict where and at which level CPT-violating 
effects may show up: this is an experimentally-driven search, 
where physical systems with the highest sensitivity, such as the 
neutral kaon system, play the leading role. 

One of the most significant tests of CPT invariance is the one 
obtained by means of the Bell-Steinberger (BS) relation: a relation 
which makes use of unitarity (or the conservation of probability) 
to connect a violation of CPT in the neutral kaon 
Hamiltonian ($m_{\kno} \not= m_{\knb}$ 
and/or $\Gno\not=\Gamma_{\knb}$) to the observable CP-violating 
interference of $K_L$ and $K_S$ decays into the same final state $f$. 
The key equation following from the BS relation is~\cite{Ambrosino}:
\beq
     \left[{\frac{\GAMS+\GAML}{\GAMS-\GAML}}+i\tan\phi_{\rm SW}\right]
      \left[ \frac{\Re(\epsilon)}{1+|\epsilon|^2 }-i\Im(\delta) \right] = 
      {\frac{1}{\GAMS-\GAML}} \sum_f \cA_L(f) \cA^*_S(f)~,
\label{eq:b-s}
\eeq
where the two output parameters,
$\epsilon$ (CP violating and CPT conserving) 
and $\delta$  (CPT violating) are defined by 
\be
\epsilon \pm \delta =  \frac{
    -i \Im\left( m_{12}\right) - 
            \frac{1}{2} \Im\left(\Gamma_{12}\right) \pm
    \frac{1}{2} \left[ \Mnb - \Mno -\frac{i}{2} 
    \left( \Gnb -\Gno\right) \right]
                    }{
    m_L - m_S +i(\Gamma_S - \Gamma_L)/2  
                    }~.
\ee
The advantage of the neutral kaon system is that only few decay modes 
give a significant contribution to the r.h.s.~of Eq.~(\ref{eq:b-s}), 
which can be determined from experiments with good accuracy.

\begin{figure}[t]
  \begin{center}
  \includegraphics[width=0.85\linewidth]{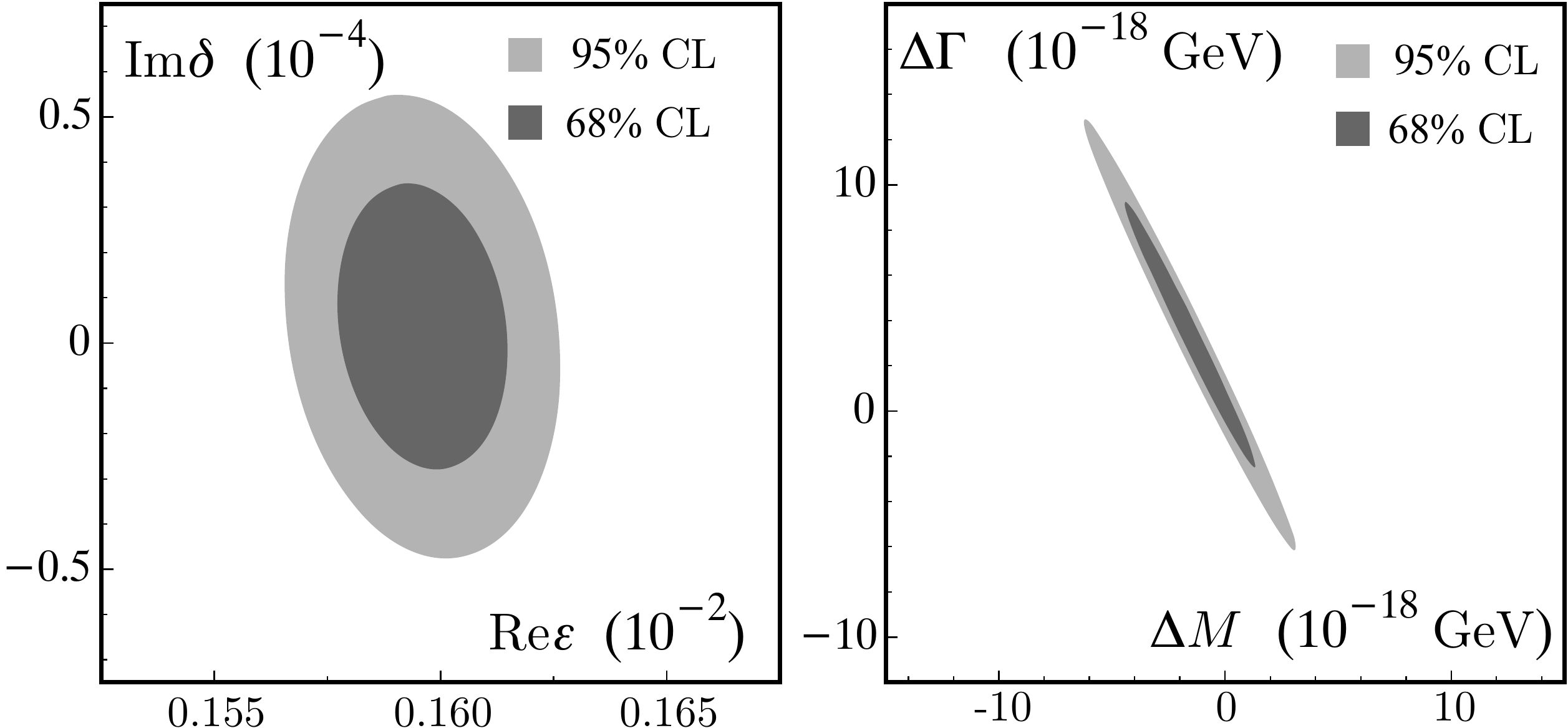}
  \end{center}
  \caption{Results of the present global fit 
to the Bell-Steinberger relation~\cite{Ambrosino}: 
allowed regions at 68\% and 95\% C.L.~in the $\Re(\epsilon)$--$\Im(\delta)$ 
plane (left) and in the $(m_{\kno} - m_{\knb})$--$(\Gno-\Gamma_{\knb})$  plane
(right).  \label{fig:CPT1} }
\end{figure}

The results of a new global fit,  including several new measurements 
by KLOE~\cite{Testa}, is shown in Fig.~\ref{fig:CPT1}. The error 
on $\Im(\delta)$ and the corresponding bound on CPT violating
effects have been improved by about a factor of two compared 
to the PDG~\cite{PDG}. As stressed by several authors, the fact 
that the difference between $\kno$ and $\knb$ masses 
is bound to be in the few$\times10^{-18}$ range (see Fig.~\ref{fig:CPT1}) 
could suggest that we are really probing Planck-scale physics. 
This fact should not be over-emphasized (there is no guarantee 
that CPT-violating effects scale linearly with $M_K/M_{\rm Planck}$), 
but it is a clear demonstration 
of the unique sensitivity of the neutral kaon system. 

The KLOE experiment is an ideal framework to make 
many other tests of CPT invariance and quantum coherence~\cite{Testa}. 
The most remarkable example is the limit set on the 
$\kno$--$\knb$ decoherence parameter, $\zeta_{0\bar0}$,
which can be bounded by the  absence of equal-time 
and equal-final-state events in the 
$\Phi \to  K^0 {\bar K}^0 \to f_1(t_1)  f_2(t_2)$ decay chain.
The recent KLOE analysis 
has set the bound $\zeta_{0\bar0} < 5 \times 10^{-6}$ 
at 95\%~C.L.~\cite{Deco}. This result 
is five order of magnitude better (!) with respect 
to the previous limit on $\zeta_{0\bar0}$ (set by CPLEAR)
and with respect to limits recently set at $B$ 
factories on  the corresponding $B$-physics 
observable~\cite{Testa}.  

\section{\boldmath  {\em The holy grail} of kaon physics: 
$K\to \pi \nu\bar\nu$ \& other ultra-rare decays \unboldmath}

Rare decays mediated by flavour-changing neutral currents (FCNCs)
are one of the best tools to investigate the flavour structure of 
physics beyond the SM. Among the many rare $K$ and $B$ decays, 
the $\kpn$ and $\klpn$ modes are unique since their SM branching ratios
can be computed to an exceptionally high degree of precision, not
matched by any other FCNC processes involving quarks. 
It is then not surprising that $K\to \pi \nu\bar\nu$  decays continue to 
raise a strong theoretical interest, both within 
and beyond the SM~\cite{Haisch,Tarantino,Smith}.

\begin{figure}[t]
  \begin{center}
   \vskip -0.4 cm
   \includegraphics[width=0.85\linewidth]{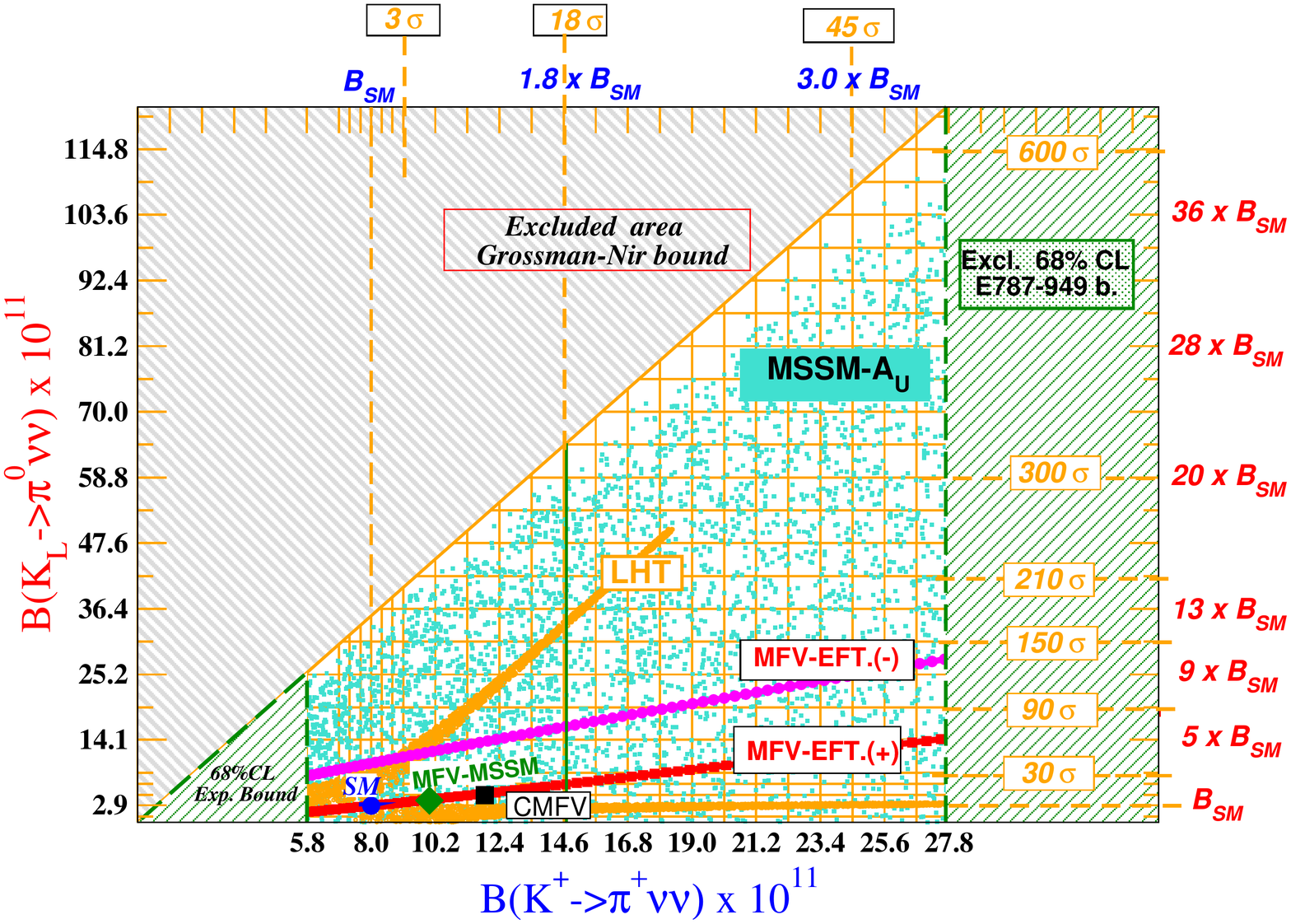}
   \vskip -0.8 cm
  \end{center}
  \caption{Predictions of different NP models for $\BR(\kpn)$
and $\BR(\kpn)$  [courtesy of F.~Mescia]. The 95\%~C.L.~exluded areas of 
$\BR(\kpn)$ refer to the result of the BNL-E787/949 
experiment~\cite{E949} \label{fig:Mesciaplot}}
\end{figure}

Beside several NP analyses~\cite{Tarantino,Smith}, 
which confirm the high discovery potential of these channels
(see Fig.~\ref{fig:Mesciaplot}),
it is worth stressing the three important improvements 
on the SM predictions of $K\to \pi \nu\bar\nu$ rates 
since KAON 2005: i) the NNLO calculation 
of the dimension-six charm-quark contribution to $\kpn$~\cite{NNLL}; 
ii) the first complete analysis of dimension-eight and long-distance 
(up-quark) contributions relevant to $\kpn$~\cite{IMS}; 
iii) a new comprehensive analysis of matrix-elements and isospin-breaking effects,
relevant to both channels~\cite{MS}.
Thanks to these recent works, the irreducible theoretical 
uncertainties on both branching ratios are at the few \%
level. 
The present SM predictions are~\cite{Haisch} 
\bea
\BR(K^+\to\pi^+\nu\bar\nu)_{\rm SM} & = (2.54 \pm 0.35) \times 10^{-11} \, , \\
\BR(K_L\to\pi^0\nu\bar\nu)_{\rm SM} & = (7.96 \pm 0.86) \times 10^{-11} \, , 
\eea
where a good fraction of the error (especially in the $K_L$ case) 
is the parametric uncertainty induced by $V_{ts}$ 
and $V_{td}$, whose knowledge will certainly improve 
in the near future.

The slow experimental progress on these channels 
is a big shame of the particle-physics community 
(and, especially, of its funding system):
several interesting proposal (even successfully running 
experiments) have been stopped or suspended for reasons which have nothing 
to do with physics. Fortunately this has not discouraged everybody!
The P326 experiment at CERN~\cite{Ruggiero} and the 
possibility to move and improve the E391a experiment~\cite{E391a} 
at the JPARC accelerator~\cite{JPARC} are very promising 
prospects for the charged and the neutral channel, respectively. 
These challenging experiments, which aim at $\cO(10\%)$ measurements 
of the branching ratios, should receive the strongest 
support of our community. Their results will be 
very interesting also in the LHC era: they will alllow us 
to investigate the flavour structure of physics beyond the 
SM~\cite{FlavLHC}. Needless to say that further 
proposals (or upgrades) would also be extremely welcome,
since neither P326 nor the JPARC experiment will be able 
to fully exploit the discovery potential of these rare modes.

\section{Conclusions}
 
This conference has clearly demonstrated the 
high interest of the recent results obtained in kaon physics, 
and the promising future of this field. As I tried to outline, 
kaon physics is still a privileged observatory for 
improving the determination of fundamental  SM parameters (such as $V_{us}$ and $m_{u,d,s}$);
performing high-precision tests of weak-current universality
 (with important implications on beyond-the-SM phenomenology);
understanding in a deeper way the structure of the QCD vacuum;
testing the basic assumptions of quantum field theory; 
shedding light on the flavour structure of physics beyond the SM. 

We learned a lot from kaon physics, but exciting 
lessons could still be ahead of us~\ldots 

\section*{Acknowledgment}
\noindent
I wish to thank the organizers of this conference, and especially
Caterina Bloise, for having created a very stimulating atmosphere.
A special thank is due to Paolo Franzini, who proposed me for 
this summary: now that is over, I should admit that it has been 
a very instructive experience!

\end{document}